\documentstyle[epsfig,prl,aps]{revtex}
\def\simleq{\; \raise0.3ex\hbox{$<$\kern-0.75em \raise-1.1ex\hbox{$\sim$}}\; }
\def\simgeq{\; \raise0.3ex\hbox{$>$\kern-0.75em \raise-1.1ex\hbox{$\sim$}}\; }
\newcommand{\be}{\begin{eqnarray}}
\newcommand{\ee}{\end{eqnarray}}

\begin{document}
\twocolumn[\hsize\textwidth\columnwidth\hsize\csname @twocolumnfalse\endcsname
%[\hsize\textwidth\columnwidth\hsize\csname@twocolumnfalse\endcsname
\hskip15.cm{SNS-PH-01-11}
\title{Generation of Magnetic Fields and Gravitational
Waves at Neutrino Decoupling}
%\end{center}
\author{Alexander D. Dolgov$^1$ and Dario Grasso$^2$}
\address{$^1$I.N.F.N., Sezione di Ferrara, Via del Paradiso 12, 
44100 Ferrara, Italy\\
and ITEP, Bol.Cheremushkinskaya, 25, 117259, Moscow, Russia\\
$^2$ Scuola Normale Superiore, P.zza dei Cavalieri 7, 56126 Pisa, Italy\\
and I.N.F.N. Sezione di Pisa, Italy}
\maketitle
\begin{abstract}
We show that an inhomogeneous cosmological lepton number may have 
produced turbulence in the primordial plasma when neutrinos entered 
(almost) free-streaming regime. This effect may be responsible for 
the origin of cosmic magnetic fields and give rise to a detectable 
background of gravitational waves. An existence of inhomogeneous
lepton asymmetry could be naturally generated by active-sterile
neutrino oscillations or by some versions of Affleck and Dine
baryogenesis scenario. 
\end{abstract}
%
%\vskip0.3cm
%\vskip-0.5cm
\vskip1pc]
The idea that the early universe went through
one or more turbulent phases is recurrent in the scientific
literature since the beginning of the modern cosmology. 
Among other reasons, cosmic turbulence 
was often invoked to explain the origin of the magnetic
fields (MFs) observed in most spiral galaxies and galaxy clusters
\cite{Kronberg,report}. One of the first attempts \cite{Harrison} was based 
on the observation that in the radiation
era a weak MF should be generated by turbulent eddies because the
rotational velocities of relativistic electrons and non-relativistic
ions would change differently during universe expansion. 
Subsequently, turbulence may have induced a Magneto-Hydro-Dynamical (MHD) 
dynamo which amplified exponentially the field until equipartition between 
the plasma turbulent kinetic energy and the MF energy is eventually reached.
This nice scenario, however, remains unaccomplished in the absence
of a plausible mechanism of turbulence generation.
Since gravitational forces conserve angular moment, vorticity must
have been produced by non-gravitational forces. 
Primordial phase transitions might have produced MFs if they were
first order \cite{Hogan}. The idea is that the expanding
bubble walls between two phases give rise to small electric
currents which power a seed MF. Turbulence appears when bubbles
collide at the end of the phase transition, or hydrodynamic
instabilities develop on the bubble walls, producing MHD
amplification of the seed field. Several applications of this idea
have been studied for the case of the quark-hadron (QHPT)
and electroweak (EWPT) phase-transitions (see
\cite{report} and refs. therein). Unfortunately, it is still
unclear if any of these transitions can really be first
order. Furthermore, the major problem with this kind of scenario,
is that it can hardly account for large scale MFs. The reason is
that the comoving coherence length of the MF is at most given
by the Hubble horizon at the phase transition 
which is much smaller than a typical galaxy size. Although the
coherence length may grow due to MHD effects  this generally happens
to the expenses of the MF strength. Careful studies showed that 
the EWPT and QHPT cannot account for the galactic and cluster MFs
\cite{Son}. To overcome the small scale problem
several mechanisms of generation of seed magnetic fields at 
inflationary stage have been proposed which are operative if
conformal invariance of electromagnetic interaction is 
broken~\cite{turner,dolgov93} (for more references
see the review~\cite{report}). 

In this Letter we propose a new mechanism for the generation of
cosmic MFs. In contrast to many previously considered
mechanisms it does not demand any new strong physical assumptions
and can be realized, in particular, in almost minimal standard
model of particle physics. It might operate at neutrino decoupling
epoch, $T\sim 1-2$ MeV,
i.e. when the Hubble horizon was considerably larger than what it
was at EWPT and at QHPT. Our basic assumption is that the net lepton
number density ($N_a({\bf x}) \equiv n_{\nu_a}({\bf x}) - n_{\bar
\nu_a}({\bf x})$,\ \ $a = e,\mu,\tau$) of one, or more, neutrino
species was not uniform before neutrino decoupling and changed in space
over some characteristic scale $\lambda$ which could be smaller than
the Hubble horizon at the decoupling time. 

As a result, when the neutrino mean free path $\ell_\nu(T)$ grew and 
became comparable to $\lambda$, neutrino currents should be
developed along the density gradients. We will show that
elastic scattering of the diffusing neutrinos on
electrons and positrons would be able to accelerate the
electron-photon fluid producing vorticity in the plasma.
Turbulence may have developed by this process in the short
interval of time during which the random forces due to neutrino
elastic scattering overcome the shear viscosity force. Depending
on the amplitude and wavelength of the fluctuations of $N_a({\bf x})$,
this period could be sufficient for the MHD engine to generate magnetic
field in equipartition with the turbulent kinetic energy. The seed
field required to initiate the process arises naturally as a consequence 
of the difference between the $\nu_a e^-$ and $\nu_a
e^+$ cross sections and of the neutrino-antineutrino local
asymmetry. Furthermore, we will show that turbulence should give
rise to a background of gravitational waves.

The first step of our computation is to determine the neutrino
momentum flux produced by the inhomogeneous $N_a({\bf x})$. 
Starting form the Boltzmann equation we obtain the equation
describing evolution of the average flux of $i$-th component of
neutrino momentum
\begin{equation}
\frac{\partial}{\partial t} {K}_i({\bf x},t)+4 H\,K_i({\bf x},t)
+\frac{\partial}{\partial x^j} {K}_{ij}({\bf x},t)
= -\tau_w^{-1} K_i
\label{diffeq}
\end{equation}
where 
\be
{K}_i = \int k_i~f_\nu (E,{\bf k}) \frac{d^3{\bf k}}{(2\pi)^3},
\label{ki} ~~~~ {K}_{ij} = \int {k_i k_j \over E}~f_{\nu}(E,{\bf k}) 
\frac{d^3{\bf k}}{(2\pi)^3} \nonumber
\ee
In the above $E$ and ${\bf k}$ are respectively the neutrino energy and
spatial momentum, $H$ is the universe expansion rate, $\tau_w$ is the 
effective weak interaction time, and  $f_{\nu}(E,{\bf k},{\bf x}, t)$ is
the neutrino distribution function.
As we see in what follows, the source of the magnetic field
is proportional to curl of electric current, ${\bf \nabla}\times {\bf J}$, 
which in turn is proportional to the local vorticity 
of the source term $\partial_i K_{ij}$ in Eq.(\ref{diffeq}). The latter is 
nonvanishing for anisotropic random initial distribution of neutrino leptonic
charge and is numerically close to $K_{ij}$ divided by $\lambda$. 
Hence, Eq.(\ref{diffeq}) becomes 
\begin{eqnarray}
\label{diffeq2}
\frac{\partial}{\partial t} {\mathbf{\cal K}}_\lambda ({\bf x},t) 
&\simeq & \frac { {\hat V}_\lambda^{in}}{\lambda} 
\left(\frac {\delta n_\nu}{n_\nu}\right)_\lambda
\exp\left(- \int_0^t \frac{\ell_\nu(t')}{\lambda^2}~dt'\right)\\
&-& 4 H {\mathbf {\cal K}}_\lambda ({\bf x},t) - 
\tau_w^{-1} {\bf{\cal K}}_\lambda({\bf x},t)
~,\nonumber
\end{eqnarray}
where ${\mathbf{\cal K}}({\bf x},t) \equiv {\bf K}/\rho_\nu$ is 
the specific momentum flux on the length-scale $\lambda$, 
and ${\hat V}_\lambda^{in}$ is a unit vector parallel to the 
initial value of the vector $\partial_j K_{ij}$, which has a non-zero 
vorticity, i.e. ${\bf \nabla}\times {\hat V} \neq 0$.
The exponential in the first term of the r.h.s.
of Eq.(\ref{diffeq2}) has been inserted to account for the
damping of the fluctuations due to neutrino diffusion. The
last term in the r.h.s. of Eq.(\ref{diffeq2}) represents the force
per unit mass between the drifting neutrinos and the
electron-positron fluid due to their scattering. It is understood that
$\lambda$ changes with time due to universe expansion.
Since at $T \simeq 1$ MeV, the Compton scattering rate was much
larger than the universe expansion rate $H$, electrons,
positrons and photons formed a (almost) single relativistic fluid at
neutrino decoupling time (the role of baryons was negligible 
at this epoch). We assume the electron-photon fluid was
approximately homogeneous before neutrino decoupling. This is
possible if the local excess of neutrinos of a given
species is balanced by a deficiency of neutrinos of a different
species (also sterile). The Euler equation of the fluid is
\begin{equation}
\frac{\partial {\bf v}}{\partial t} \simeq \tau_{\nu e}^{-1} \frac{ \rho_\nu}
{(\rho + p)\gamma^2}\left({{\mathbf {\cal K}}}_{\nu} + 
{{\mathbf {\cal K}}}_{{\bar \nu}}\right) - H {\bf v} + 
\eta  \nabla^2 {\bf v} ~, 
\label{eqmot}
\end{equation}
where $\rho$ and $p$ are the energy density and pressure of the
electron-photon fluid, $\gamma = \sqrt{1 - v^2}$ is the Lorentz
factor,  $\tau_{\nu e} \equiv \tau_{\nu e^-} + \tau_{\nu e^+}$
is the neutrino-electron(positron) mean collision time,
and $\eta =  4\rho_\nu \ell_\nu /15(\rho + p)$
is the shear viscosity due to neutrino diffusion. This 
approximation for the viscosity is valid in the limit when the
neutrino mean free path is smaller than the characteristic scale
of the problem. In our case this is not always true and the 
viscosity can be smaller.
In Fig.1 the macroscopic velocity $v$ of the electron-photon fluid 
(continuous line) is presented as a function of the ratio 
$x \equiv t/t_d$ ($t_d$ is the neutrino decoupling time). It is
obtained by numerical solution of the coupled set of 
Eqs.~ (\ref{diffeq2}) and (\ref{eqmot}) for a particular choice of 
$\delta n_\nu$ and characteristic size of the fluctuations
(see the figure caption). 
It is visible from this figure that the velocity
grows rapidly when the neutrino mean free path becomes comparable
to the fluctuations size $\lambda$ and it is suppressed soon later
when the viscosity on that length scale becomes dominant.
\vskip-0.5cm
\begin{figure}
\centerline{\protect\hbox{
\psfig{file=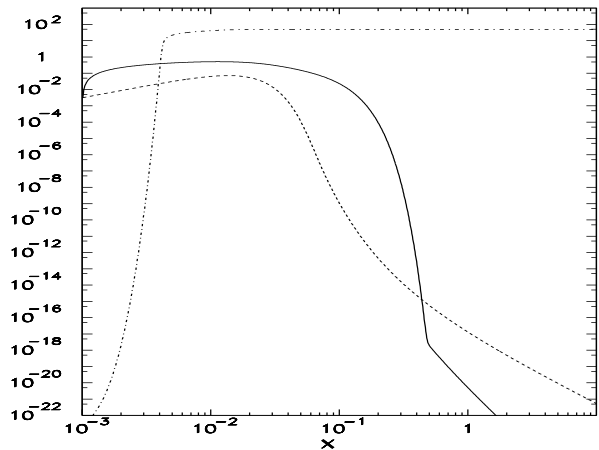,height=7.3cm,width=9.3cm,angle=0}}}
\caption{The local fluid velocity $v$ (continuous line), the
specific neutrino momentum flux ${\bf{\cal K}}$ (dashed line), and
the ratio $b = \vert e\vert B/T^2$ (dotted-dashed line)
as functions of the time parameter $x$. We assumed
here $\delta n_\nu/n_\nu = 1$ over a comoving length scale
$\lambda_d \equiv (\lambda/H^{-1})_{t=t_d} = 10^{-2}$.
The quantities plotted here represent avarages on the same scale.}
\end{figure}

In this Letter we do not investigate the velocity and MF power spectrum. 
We only observe here that due to the rapid grow of the
viscosity with time, the turbulence spectrum may be quite
different from the Kolmogoroff's one.

We now turn to the problem of magnetic field generation. Starting from
Maxwell equations in the primordial plasma and neglecting terms 
proportional to the inverse of the electric conductivity we obtain after 
simple algebra
\begin{equation}
\label{magnetic}
\partial_t {\bf B} + 2H{\bf B} = 
{\bf \nabla} \times \left( {\bf v}\times {\bf B}\right)
+ \kappa^{-1}  {\bf  \nabla} \times {\bf J_{ext}}
\label{dtB}
\end{equation}
where $\kappa \sim T/\alpha$ \cite{turner,BayHei}
is the electric conductivity of the 
relativistic cosmological plasma ($\alpha$ is the fine structure constant), 
and $2HB$ is the damping term related to the cosmological
expansion.  The validity of our approximation can be  
easily verified by noting that the magnetic diffusion length
$\displaystyle L_{\rm diff}(t) = \sqrt{t/4\pi \kappa}$ exceeds safely the
characteristic length of the relevant process, namely  
the neutrino mean-free-path $\ell_\nu(t) \sim 1~{\rm s}
~(t/t_d)^{5/2}$. Using the expression of the electric conductivity
for relativistic plasma presented above we find
that the condition $L_{\rm diff}(t) \gg \ell_\nu(t)$ is
comfortably fulfilled at any time essential for the presented
mechanism. We also checked that relativistic corrections to 
Eq.(\ref{magnetic}) do not affect significantly our results even if 
$v \simeq 1$.
Thanks to the absence of magnetic diffusion, the first term 
in the r.h.s. of Eq.(\ref{magnetic}) ensures an exponential
amplification of any preexistent seed field which is accounted by the
the term $\kappa^{-1}  {\bf  \nabla} \times {\bf J_{ext}}$ in Eq.(
\ref{magnetic}). Interestingly, neutrino diffusion in the
presence of a inhomogeneous neutrino-antineutrino asymmetry
naturally provides such a seed. The relevant point here is that,
due to charge symmetry breaking in the standard model, the
neutrino-electron and the neutrino-positron weak cross
sections are different. As a consequence electrons and positrons
are subject to different forces in the presence of a net flux of
neutrinos and this gives rise to a non-zero electric current.
The latter can be estimated as follows. The equation of motion
of electron (or positron) in plasma is $\dot p_e = \Phi_\nu \sigma_{\nu e} 
\Delta p - p_e /\tau_e$
where $\Phi_\nu = \delta n_\nu$ is the neutrino flux with $\delta n_\nu$
being an excess of neutrino number density, $\Delta p \sim T$ is the
transferred momentum, and 
$\tau_e = (\sigma_T n_\gamma)^{-1}$ is the characteristic damping time
due to $e^{\pm}$ interactions with photons, 
$\sigma_T = 8\pi\alpha^2 /3m_e^2$ is the Thomson cross-section and
$n_\gamma = 0.24 T^3$ is the photon number density. 

Since the $\nu e^-$ and $\nu e^+$ cross-sections are different,
$\sigma_{\nu e^-} = G_F^2 s /\pi \approx 10 G^2_F T^2$ and
$\sigma_{\nu e^+} = G_F^2 s /3\pi \approx 3 G^2_F T^2$ (barring
finite electron mass corrections), the drift
velocities of electrons and positrons would be different and a
non-zero electric current would be induced by neutrino flux. Using
the previous expressions the difference of $(e^--e^+)$-velocities can be 
estimated as $\Delta v \approx 2\cdot 10^{-19} (\delta n_\nu /n_\nu)
(T/{\rm MeV})^3$.
Correspondingly the electric current induced by
neutrinos is given by:
\be
J_{\rm ext} = 4\cdot 10^{-20}e T^3 \left({T\over {\rm MeV}}\right)^3
\left({\delta n_\nu \over n_\nu}\right)_\lambda~, 
\label{Jext}
\ee
which corresponds to a seed field with strength 
$B^{\rm seed}_\lambda \approx 10^{-22}  
\left( T\over {\rm MeV}\right)^2$ at a time  $t/\lambda \sim 1$.
%Before coming to the numerical results it is useful to comment on  
%the three basic ingredients required for the generation of  MFs at neutrino
%decoupling, namely: ~1) an {\bf out-of-equilibrium} condition is required
%for a non-vanishing neutrino, or antineutrino, momentum flux to
%develop (see Eq.(\ref{diffeq})); ~2) {\bf charge parity violation} 
%is needed to have different interaction rates for charge carriers of
%the opposite sign; ~3) finally, an {\bf inhomogeneous lepton asymmetry}
%has to be invoked in order for the effect produced by neutrinos be
%not erased by antineutrinos. The first two ingredients
%are naturally provided by standard particle physics and cosmology.
%The third ingredient we assume to be supplied by moderately new 
%physics which might took place just before neutrino decoupling or 
%during those processes responsible for the observed baryon asymmetry 
%of the universe \cite{Dolgov}.
We are now able to solve numerically Eq.(\ref{magnetic}) by
coupling this equation to Eqs.(\ref{diffeq2},\ref{eqmot}). 
Instead of the absolute value of ${\bf B}$ it is
convenient to focus on the ratio $b \equiv \vert e B\vert/T^2$
which is a constant quantity for a frozen-in MF in the absence of
entropy production.  From Fig.1 it is evident at a glance the huge 
amplification of $b$ which takes place in the short interval of time during 
which fluid ``turbulence" is active. A suitable cutoff has been used in our
computation to account for the saturation of the amplification
process which has to come in when the MF approaches energy
equipartition with the fluid motion, i.e. when $\displaystyle
{B^2}/{4\pi} \sim \left(\rho + p\right) v^2$. We think that a
more elaborate treatment of non-linear MHD effects might only
slightly change the knee of $b(x)$ without affecting significantly
our main results. The final value of $b$ depends on the rapidity
of the amplification process. Since $\displaystyle b(x) \propto
\exp\left(v(x)\,x /\lambda \right)$, it is clear that
equipartition will be reached more quickly, i.e. when $v$ is yet 
not suppressed by viscosity, for fluctuations having large
amplitude and small sizes. In Fig.2 we present the final
value of $b$ as a function of $\lambda_d$ for several values of
$(\delta n_\nu/n_\nu)$. In principle the requirement of a
successful BBN put an upper limit on the strength of the MFs,
which is roughly $b \simleq 0.1$ \cite{BBN}. This
bound, however, was obtained under the assumption that the MF was
uniform over the Hubble volume at BBN time and, depending on the details of
the MF power spectrum, it may not apply to our case. 
\vskip-0.5cm
\begin{figure}
\centerline{\protect\hbox{
\psfig{file=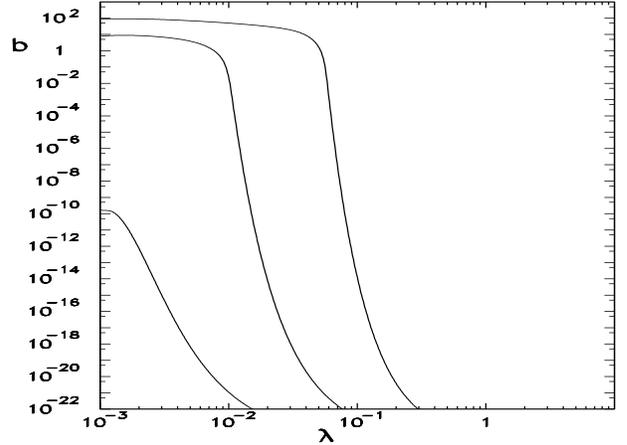,height=7.3cm,width=9.3cm,angle=0}}}
\caption{The final ratio $b = \vert e B \vert/T^2$ on the  
comoving length scale $\lambda_d$ for three values of the neutrino number 
density contrast: starting from the right curve 
$(\delta n_\nu/n_\nu)_{\lambda_d} = 1,~10^{-1},~10^{-2}$.}
\end{figure}

In principle inhomogeneous MFs can be subject to dissipation due
to their back-reaction on the plasma. In fact, tangled MFs give 
rise to an anisotropic pressure that could induce
plasma oscillations (mainly Alfv\'en waves) about a force-free 
configuration.
These oscillations might be rapidly damped because of the high fluid
viscosity which implies a dissipation of magnetic energy \cite{Olinto}. 
In our case, it is easy to verify that during the neutrino 
decoupling process the oscillation frequency of Alfv\'en waves is always 
smaller than the characteristic rate at which MFs  are produced which is 
$\simleq H$ so that these MHD modes are not excited.  
MF dissipation could also take at photon 
decoupling. However, it was showed in Ref.\cite{Olinto} that such an 
effect is negligible for MFs extending on galactic scale if
their present time strength is smaller than $10^{-9}$ Gauss. 

The generation mechanism that we have discussed 
gives rise to MFs of intensity $B_0 =
b\  8 \times 10^{-6}$ Gauss at the present time with a coherence length
$\lambda_0 = \lambda_d~ r_H(t_d)~ (T_d/T_0) \simeq 10^2
~\lambda_d$ pc. Galactic MFs are observed with characteristic
strength of the order of $1~\mu G$ extending over scales $\sim 1$
kpc \cite{Kronberg}. Taking into account flux conservation during
the protogalaxy collapse, the primordial origin of galactic fields
would require a protogalactic field with the strength $\sim 10^{-10}$
Gauss and the coherence length of 0.1 Mpc \cite{Kronberg,report}.
Although this scale is much larger than the coherence
length predicted by our model, it is natural to expect that some
homogenization could take place during galaxy formation. Since the
field orientation is random over scales larger than $\lambda_0$,
the predicted mean field on the protogalactic scale will be
obtained by a suitable volume average \cite{Hogan}
\begin{equation}\label{prediction}
  B(0.1~{\rm Mpc}) \simeq B_0 \left(\frac {\lambda_0}{0.1~{\rm
  Mpc}}\right)^{3/2} \simeq 10^{-10}b~{\lambda_d}^{3/2}~{\rm G}.
\end{equation}
From this equation and our previous results (see Fig.2) we
find that galactic MFs may be a product of
neutrino number fluctuations with the amplitude $\sim 1$ extending over
scales comparable to the Hubble horizon at neutrino decoupling. It
is remarkable that MFs with this intensity may have produced
observable effects on the CMB  and have interesting consequences
for structure formation \cite{report}. Fluctuations 
with a smaller amplitude could still have played a role in the generation 
of galactic MFs if a galactic dynamo, or even much less efficient 
amplification  processes, took place during or after galaxy formation.

Another interesting consequence of stirring the primordial plasma
by neutrino inhomogeneous diffusion is the production of
gravitational waves (GW). The generation of a cosmic background of
GW by turbulence produced at the end of a
first-order phase transition was discussed in Ref.\cite{KKT}. In
our case GW with the largest amplitude are produced with a
comoving wavelength $\lambda_0 = \lambda_d~10^2$ pc which
corresponds to the present time frequency
$  \omega_0 \simeq 10^{-9} \lambda_d^{-1}\ {\rm Hz}$.
The GW production took place when $\ell_\nu(x_*) = \lambda(x_*)$,
i.e. $x_* = \sqrt{\lambda_d}$, and it lasted for a time interval
comparable to $H(x_*)$. Following ref.\cite{KKT} we estimate
the energy density parameter of GW with frequency $\omega_0$ to be
\begin{eqnarray}
  \Omega_{\rm GW} h^2 &\simeq& 10^{-5}~H(x_*)~\lambda(x_*) v^6(x_*)
  \nonumber \\
  &\simeq& 3\times 10^{-5} 
  \lambda_d^{3/4}~ v^6(x_*)~.
  \label{omegagw}
\end{eqnarray}
From our previous results it follows that a GW
background produced by neutrino number fluctuations of amplitude
$\simleq 1$  with $\lambda_d < 10^{-3}$ may be
detectable by future GW space based observatories \cite{Maggiore}.
As tangled MFs can also act as a source of GW \cite{Durrer}, a
further, and perhaps dominant, contribution to the GW background
may come from MFs produced during the decoupling process.

We conclude this Letter by observing that large neutrino number
density fluctuations, as those required to power the effects
discussed above, might be generated by several mechanisms.
One possible way to generate inhomogeneous and large lepton asymmetry
at large scales together with a small baryon asymmetry could be 
achieved~\cite{Dolgov,dolkir} in the frameworks of
Affleck-Dine~\cite{affleck} baryo(lepto)-genesis scenario. Another
very interesting mechanism based on active-sterile neutrino
oscillations was proposed recently by DiBari
\cite{DiBari}.
Such a mechanism naturally gives rise to domains
where neutrinos, or antineutrinos, of a given species are
strongly converted into sterile neutrinos (or sterile antineutrinos),
hence $\delta n_\nu/n_\nu \sim 1$. The typical domain
size is determined by the neutrino mean free path at the critical
temperature at which neutrino conversion takes place
$\displaystyle T_c \simeq 15~{\rm MeV}\left(\vert \delta
m^2\vert/{\rm eV}^2\right)^{1/6}$, which implies $\displaystyle
\lambda_d \simeq 10^{-3}~\left( \vert \delta m^2\vert/{\rm
eV}^2\right)^{-2/3}$. A nice feature of this scenario is that
it practically does not need to invoke new physics, except for
only one that there exists a sterile neutrino mixed with an
active one. In this case very small inhomogeneities in the baryon
number density, which is known to exist in the early universe, would
give rise to a very strong amplification of initially negligible
lepton asymmetry. Since the initial excess of energy density of active 
neutrinos is exactly compensated by the deficit in sterile ones 
such a model does not suffer from a possible distortion of approximate
isotropy of the cosmic microwave background radiation.
Both mechanisms discussed here produce isocurvature
fluctuations in the neutrino fluid. Chaotic vector and tensor
perturbations, which may be absent initially, are produced when neutrino 
start to diffuse according to the mechanism discussed in this 
Letter.
\acknowledgments

The authors wish to thank P.~Di Bari and A.~Rossi for useful 
discussions. 
D.G. thanks the Physics Department of Ferrara University for the kind 
hospitality and financial support during part of the preparation of 
this work.


\vskip-0.5cm
\begin{thebibliography}{99}

\bibitem{Kronberg} P.P.~Kronberg, Rep.~Prog.~Phys. {\bf 57}, 325
(1994).

\bibitem{report} D.~Grasso and H.R.~Rubinstein, Phys.\ Rept.\ 
{\bf 348}, 161 (2001).

\bibitem{Harrison} E.R.~Harrison, Mon.~Not.~Roy.~Astron.~Soc.,
{\bf 147}, 279 (1970); Phys.~Rev.~Lett. {\bf 30} 188 (1973).

\bibitem{Hogan} C.H.~Hogan, Phys.~Rev.~Lett. {\bf 51}, 1488
(1983).

\bibitem{Son} D.T.~Son, Phys.\ Rev.\ {\bf D59} 063008 (1999).

\bibitem{turner} M.S.~Turner and L.M.~Widrow, 
Phys.\ Rev.\ {\bf D37} 2743 (1988).

\bibitem{dolgov93} A.D.~Dolgov, Phys.\ Rev.\ {\bf D48} 2499 (1993).

\bibitem{BayHei} J.~Ahonen and K.~Enqvist, Phys.\ Lett.\  {\bf B382},
40 (1996); G.~Baym and H.~Heiselberg, Phys.\ Rev.\ {\bf D59}, 5254
(1997).

\bibitem{BBN} D.~Grasso and H.R.~Rubinstein, Phys.\ Lett.\  {\bf B379},
73 (1996).

\bibitem{Olinto} K.~Jedamzik, V.~Katalini\'c and A.~Olinto,
Phys.\ Rev.\ {\bf D57}, 3264 (1998); K.~Subramanian and D.~Barrow,
Phys.\ Rev.\ {\bf D58}, 083502 (1998).

\bibitem{KKT} M.~Kamionkowski, A.~Kosowsky and M.~Turner,
Phys.\ Rev.\ {\bf D49}, 2837 (1994).

\bibitem{Maggiore} M.~Maggiore, Phys.\ Rept.\ , {\bf 331}, 283
(2000).

\bibitem{Durrer} R.~Durrer, P.G.~Ferreira and T.~Kahniashvili,
Phys.\ Rev.\ {\bf D61}, 043001 (2000).

\bibitem{Dolgov} A.D.~Dolgov, Phys.\ Rept.\ {\bf 222}, 309 (1992).

\bibitem{dolkir}
A.D.~Dolgov and D.P.~Kirilova, J.~ Moscow~Phys.~Soc.~{\bf 1}, 217 (1991).

\bibitem{affleck} I.~Affleck and M.~Dine, Nucl.\ Phys.\ {\bf 249}, 361
(1985).

\bibitem{DiBari} 
P.~Di Bari, Phys.\ Lett.\  {\bf B482}, 150 (2000).

\end{thebibliography}
\end{document}